\documentclass[11pt]{article}

\usepackage[preprint]{acl}

\usepackage{times}
\usepackage{latexsym}
\usepackage{multirow}
\usepackage[most]{tcolorbox}
\usepackage{listings}
\usepackage{url}
\usepackage[T1]{fontenc}

\usepackage[utf8]{inputenc}

\usepackage{microtype}

\usepackage{inconsolata}

\usepackage{graphicx}

\usepackage{booktabs}
\usepackage{amssymb}
\usepackage{siunitx}
\title{Full-Duplex-Bench-v3: Benchmarking Tool Use for Full-Duplex Voice Agents Under Real-World Disfluency}

\author{
 \textbf{Guan-Ting Lin\textsuperscript{1},
 \textbf{Chen Chen\textsuperscript{2}},
 \textbf{Zhehuai Chen\textsuperscript{2}},
 \textbf{Hung-yi Lee\textsuperscript{1}}
}
\\
\\
 \textsuperscript{1}National Taiwan University,
 \textsuperscript{2}NVIDIA
 \\
 \small{
   \textbf{Correspondence:} \href{daniel094144@gmail.com}{daniel094144@gmail.com}
 }
}

\begin{document}
\maketitle
\begin{abstract}
We introduce \textbf{Full-Duplex-Bench-v3 (FDB-v3)}, a benchmark for evaluating spoken
language models under naturalistic speech conditions and multi-step tool use.
Unlike prior work, our dataset consists entirely of real human audio annotated
for five disfluency categories, paired with scenarios requiring chained API
calls across four task domains. We evaluate six model configurations---GPT-Realtime,
Gemini Live 2.5, Gemini Live 3.1, Grok, Ultravox v0.7, and a traditional
Cascaded pipeline (Whisper$\rightarrow$GPT-4o$\rightarrow$TTS)---across accuracy,
latency, and turn-taking dimensions. GPT-Realtime leads on Pass@1 (0.600)
and interruption avoidance (13.5\%); Gemini Live 3.1 achieves the fastest
latency (4.25~s) but the lowest turn-take rate (78.0\%);
and the Cascaded baseline, despite a perfect turn-take rate, incurs the
highest latency (10.12~s). Across all systems, self-correction handling and
multi-step reasoning under hard scenarios remain the most consistent failure
modes. Demo is available at \href{https://daniellin94144.github.io/FDB-v3-demo/}{https://daniellin94144.github.io/FDB-v3-demo/}.
\end{abstract}

\begin{figure*}[htbp]
  \centering
  \includegraphics[width=\linewidth]{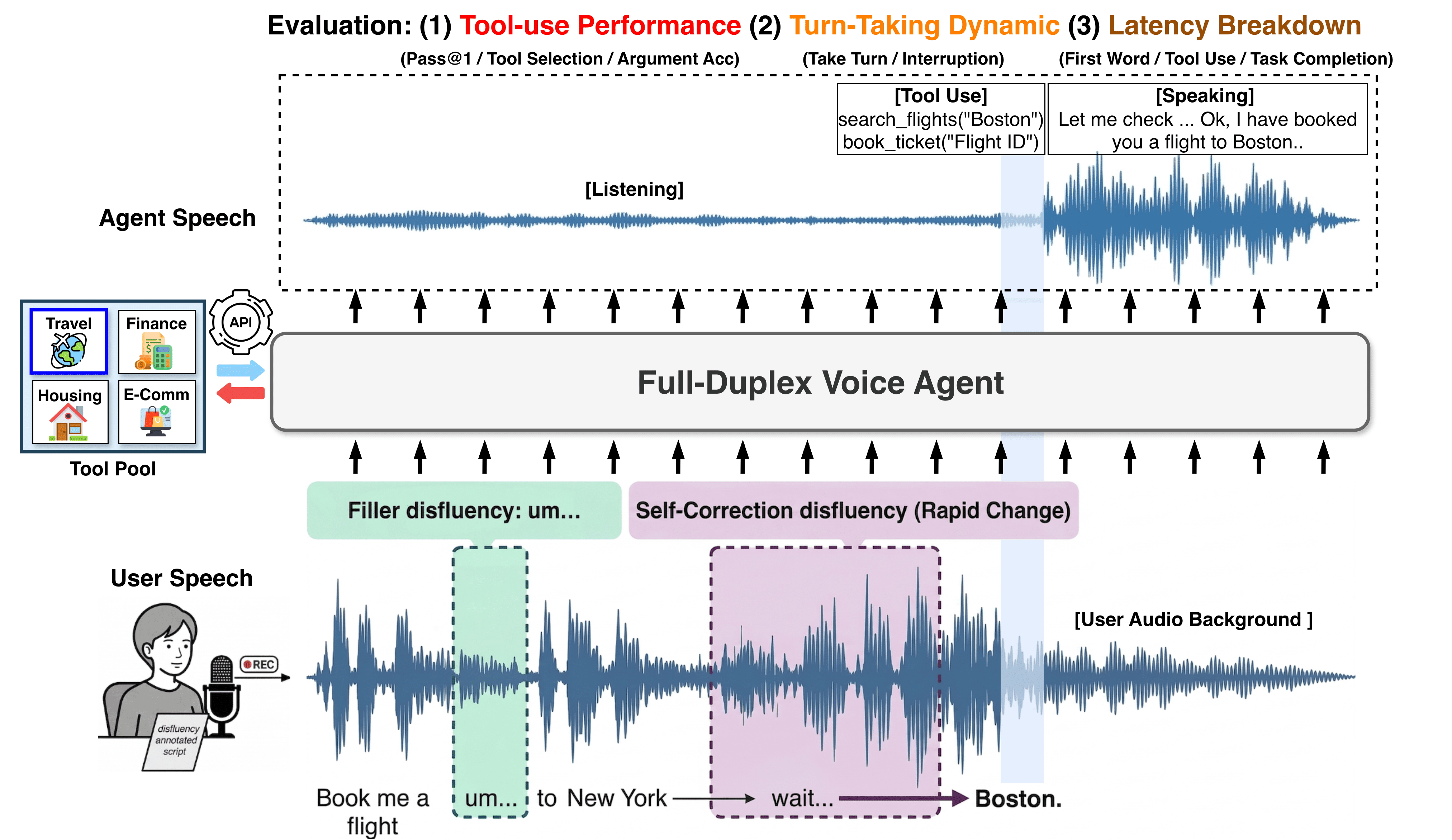} 
  \caption{The Full-Duplex-Bench-v3 framework for evaluating real-time voice agents. The diagram illustrates a full-duplex interaction where the agent continuously processes user speech containing natural dysfluencies, such as fillers and self-corrections. Full-Duplex-Bench-v3 comprehensively assesses the agent across three primary dimensions: (1) Tool-use Performance (accuracy of API selection and execution), (2) Turn-Taking Dynamic (handling interruptions and conversational flow), and (3) Latency Breakdown (timing from listening to tool execution and speech generation).}
  \label{fig:main_figure}
\end{figure*}

\section{Introduction}

The paradigm of tool use, which empowers AI systems to interact with external APIs and real-world environments, has transformed text-based Large Language Models (LLMs) into remarkable autonomous assistants. By leveraging this capability to orchestrate complex multi-step workflows in response to natural language instructions~\cite{toolllm, react}, text-based agents have moved beyond mere text generation. Yet extending these agentic capabilities to the voice modality has lagged behind: most spoken dialogue systems remain confined to conversational chat~\cite{moshi}, without the ability to invoke external APIs or execute actions on behalf of the user. 

This gap matters because voice agents are deployed precisely in contexts where the ability to \emph{act} creates value, such as checking flight prices, updating account settings, or tracking a parcel. Tool use is therefore not a technical nicety but a practical necessity. Voice interaction also introduces a challenge absent from text-based agents: latency. A spoken reply must arrive within a narrow window to feel natural~\cite{heldner2010pauses}, making the tension between careful reasoning and prompt response a first-class design concern. 

Despite its practical importance, combining tool use with low conversational
latency remains understudied. Earlier cascaded approaches such as
AudioGPT~\cite{audiogpt} and Speech-Copilot~\cite{speechcopilot} demonstrated
LLM-orchestrated tool use for speech and audio tasks, but their multi-stage
pipelines are designed for offline processing rather than real-time dialogue.
More recent efforts like StreamRAG~\cite{streamrag} and SHANKS~\cite{shanks}
target conversational settings, yet their training relies on synthetic data. Furthermore, because these models are not publicly accessible, their performance in real-world interactions remains unverified. In practice, the most capable systems are proprietary (OpenAI's Realtime API\footnote{https://openai.com/index/introducing-gpt-realtime/} and Google's Gemini Live\footnote{https://blog.google/innovation-and-ai/models-and-research/gemini-models/gemini-3-1-flash-live}), yet they have not been evaluated under controlled, reproducible conditions.

Existing benchmarks are ill-suited for evaluating real-world tool execution. While the Full-Duplex-Bench series~\cite{fdb1, fdb15, fdb2} pioneered evaluations for turn-taking, overlaps, and multi-turn dialogues, existing tool-use datasets still fall short. Benchmarks featuring real speech, such as Audio MultiChallenge~\cite{multichallenge} and WildSpeech-Bench~\cite{wildspeech}, lack tool-use evaluations entirely. Conversely, those that do evaluate tool use---such as $\tau$-Voice~\cite{tau},  AudioCRAG~\cite{streamrag} and VoiceAgentBench~\cite{voiceagentbench}---rely on synthetic audio or restrict tasks to single-step actions, stripping away the natural disfluencies of real spoken interaction.

Consider a user who says, \textit{``Book me a flight um... to New York---actually, wait... make that Boston.''} A robust agent must discard the earlier destination and update its internal state before issuing the booking call. While recent benchmarks like Audio MultiChallenge~\cite{multichallenge} evaluate semantic comprehension of mid-utterance speech repairs, and AgentChangeBench~\cite{agentchangebench} explores dynamic goal shifts in text-based workflows, testing this kind of \emph{programmatic state rollback} in continuous, multi-step spoken tool execution remains an open challenge. Full-Duplex-Bench-v3 (FDB-v3) bridges this gap by directly testing dynamic state updates against deterministic API constraints. Our contributions can be summarized below:
 
\begin{itemize}
    \item \textbf{Authentic human speech with systematic disfluency annotation.}
    Every query comes from real human recordings in uncontrolled environments,
    annotated across five disfluency categories---fillers, pauses, hesitations,
    false starts, and self-corrections---enabling fine-grained robustness
    analysis.
 
    \item \textbf{Self-correction and state rollback scenarios.}
    Twenty-one of our 100 scenarios test whether models can recognise a
    mid-utterance change of intent and correctly update downstream API
    parameters.
 
    \item \textbf{Multi-step function chaining across four task domains.}
    Each scenario requires a sequence of API calls spanning Travel \& Identity,
    Finance \& Billing, Housing \& Location, and E-Commerce Support, with fully
    deterministic outputs enabling automatic scoring.
\end{itemize}

 We evaluate six configurations---GPT-Realtime, Gemini Live 2.5, Gemini Live 3.1, Grok, Ultravox v0.7, and a Cascaded pipeline---across accuracy, latency, turn-taking, and disfluency robustness. GPT-Realtime leads on accuracy with the lowest interruption rate; Gemini Live 3.1 is fastest but has the lowest turn-take rate; and the Cascaded baseline guarantees engagement at the cost of the highest latency. Self-correction handling remains the hardest challenge: even GPT-Realtime succeeds on fewer than 59\% of this scenario.

\begin{table*}[]
\centering
\scriptsize
\begin{tabular}{ll}
\toprule
\textbf{Domain} & \textbf{Mock API Functions} \\
\midrule
\textbf{Travel \& Identity} & \texttt{search\_flights(destination, date)} \\
 & \texttt{book\_ticket(passenger\_name, flight\_id)} \\
 & \texttt{update\_travel\_profile(document\_type, document\_number)} \\
\addlinespace
\textbf{Finance \& Billing} & \texttt{query\_card\_benefits(card\_last\_4, category)} \\
 & \texttt{calculate\_currency\_exchange(amount, from\_currency, to\_currency)} \\
 & \texttt{modify\_autopay\_source(new\_account\_id)} \\
\addlinespace
\textbf{Housing \& Location} & \texttt{search\_apartments(max\_budget, amenities)} \\
 & \texttt{update\_search\_filter(condition, new\_value)} \\
\addlinespace
\textbf{E-Commerce Support} & \texttt{check\_order\_status(order\_id)} \\
 & \texttt{cancel\_pending\_action(action\_type)} \\
 & \texttt{process\_exchange(order\_id, new\_shipping\_address)} \\
\bottomrule
\end{tabular}
\caption{Overview of Domains and Mock API Functions}
\label{tab:mock_apis}
\end{table*}

\section{Related Works}

\subsection{Full-Duplex Evaluation} The Full-Duplex-Bench series~\cite{fdb1, fdb15, fdb2} progressively addressed real-time evaluation: v1~\cite{fdb1} pioneered turn-taking and interruption assessment; v1.5~\cite{fdb15} expanded to overlapping scenarios; and v2~\cite{fdb2} introduced multi-turn evaluations with an SLM-based examiner. However, existing datasets fall short in real-world applicability---most rely on TTS-synthesized audio rather than authentic queries. Audio MultiChallenge~\cite{multichallenge} and WildSpeech-Bench~\cite{wildspeech} incorporate real speech but lack tool-use evaluations; $\tau$-Voice~\cite{tau} and AudioCRAG~\cite{streamrag} (a spoken adaptation of CRAG~\cite{crag}) evaluate tool use but are restricted to synthetic audio or single-step calls without accounting for disfluencies. VoiceAgentBench~\cite{voiceagentbench} tests complex multi-tool workflows, its use of synthetic audio ignores the realities of human speech. Lacking real-time dynamics and natural disfluencies, it fails to measure how latency or mid-utterance corrections affect an agent. Proprietary evaluations like ComplexFuncBench Audio measure multi-step function calling but remain closed-source. Full-Duplex-Bench-v3 bridges this gap with an open, reproducible, disfluency-annotated benchmark combining real human speech with multi-step tool use.

\subsection{Tool-Use Voice Agents} Early cascaded systems such as
AudioGPT~\cite{audiogpt} and Speech-Copilot~\cite{speechcopilot} let an
LLM orchestrate external speech and audio models, but their multi-stage
pipelines preclude real-time interaction. Nowadays, the most capable tool-use real-time voice agents---GPT-Realtime, Gemini Live, Grok---are proprietary and have not been assessed under reproducible conditions. Ultravox is a notable open-weight exception, fusing a pretrained backbone with a Whisper encoder to enable native tool use without a separate ASR stage.
On the academic side, \citet{enhancing} retrieve knowledge directly from speech (RAG rather than action-oriented tool use); StreamRAG~\cite{streamrag} predicts queries in parallel with incoming speech to reduce latency; and SHANKS~\cite{shanks} uses unspoken chain-of-thought for mid-turn tool execution. However, most studies withhold complete models and inference code, and the scarcity of open-source systems hinders evaluation of complex multi-step tool execution. Full-Duplex-Bench-v3 addresses this by providing a reproducible benchmark for fair comparison across models.

\section{Benchmark Design}

We describe the benchmark construction: API design, scenario formulation with difficulty tiers, and naturalistic audio collection from human speakers.

\subsection{Task Domains and Mock APIs}

Rather than routing queries to live web services, we use locally executed mock APIs with deterministic, zero-latency responses. This isolates model reasoning and parameter-passing from confounds such as network variability or service downtime, and ensures that all measured latency reflects strictly the model's processing overhead. All expected outputs are fully deterministic, enabling automatic scoring. The benchmark spans four task domains, each with a small set of callable tools (Table~\ref{tab:mock_apis}).

\subsection{Dataset Preparation and Difficulty Tiers}

Scenarios are divided into three difficulty tiers by the number of required tool calls and reasoning complexity: \textit{Easy} (single-step), \textit{Medium} (two-step with moderate ambiguity), and \textit{Hard} (multi-step with conflicting constraints). All audio was collected from human speakers in uncontrolled environments.

Each recording is annotated for five disfluency categories, each targeting a distinct failure mode: \textit{false starts} (abandoning an intent for a new one) test whether models discard obsolete context without hallucinating tool calls; \textit{self-corrections} (updating parameters mid-sentence) assess dynamic state rollback; \textit{fillers} (e.g., \textit{um}, \textit{uh}) probe whether redundant tokens degrade accuracy or inflate latency; \textit{pauses} (mid-utterance silences) and \textit{hesitations} (filler--repetition combinations) test end-of-turn detection robustness.

\subsection{Audio Data Collection and Demographics}

The dataset comprises 100 recordings from 12 speakers, including native and non-native English speakers (Korean and Russian backgrounds) with varying accent strengths. Speakers were given detailed scenario contexts and asked to perform prompts organically. Audio was captured with everyday built-in microphones (11 of 12 setups) in environments ranging from quiet rooms to settings with mild background noise, ensuring evaluation under realistic conditions.

For trailing silence, we capture 30 seconds of each speaker's actual ambient environment rather than appending digital silence. This keeps the acoustic background coherent, closely mimicking real-world streaming interactions.

Each speaker was assigned 10 scenarios distributed across all four domains, with disfluencies proportionally represented. Twenty-one scenarios specifically feature self-correction events to test real-time state rollback. All recordings were reviewed for quality.

\section{Experimental Setup}

We evaluate six model configurations: \textbf{GPT-Realtime}\footnote{https://developers.openai.com/api/docs/models/gpt-realtime-1.5}, \textbf{Gemini Live 2.5}\footnote{https://ai.google.dev/gemini-api/docs/models/gemini-2.5-flash-native-audio-preview-12-2025}, \textbf{Gemini Live 3.1}\footnote{https://ai.google.dev/gemini-api/docs/models/gemini-3.1-flash-live-preview}, \textbf{Grok}\footnote{https://x.ai/news/grok-voice-agent-api}, \textbf{Ultravox v0.7}\footnote{https://github.com/fixie-ai/ultravox}, and a \textbf{Cascaded} baseline. All six are deployed through the LiveKit Realtime Voice Agent framework for streaming audio and real-time tool use. The first five are end-to-end speech-to-speech models; the Cascaded system follows a traditional modular pipeline---OpenAI Whisper for speech recognition, GPT-4o for reasoning and tool use, and OpenAI TTS for speech synthesis---serving as a reference point for the conventional architecture. Each model receives the same audio input.

\begin{table*}[htbp]
\centering
\resizebox{\textwidth}{!}{%
\begin{tabular}{lccccccccc}
\toprule
& \multicolumn{4}{c}{\textbf{Tool Use}} & \multicolumn{4}{c}{\textbf{Turn-Taking Dynamics}} \\
\cmidrule(lr){2-5} \cmidrule(lr){6-9}
& \textbf{Tool Sel$\uparrow$} & \textbf{Arg Acc$\uparrow$} & \textbf{Resp Qual$\uparrow$} & \textbf{Pass@1$\uparrow$} & \textbf{Take-turn$\uparrow$} & \textbf{Latency$\downarrow$} & \textbf{Interrupt$\downarrow$} & \textbf{Filler$\downarrow$} \\
\midrule
\textbf{GPT-Realtime} & \textbf{0.876} & \textbf{0.680} & \textbf{0.792} & \textbf{0.600} & 96.0\% & 6.89s & \textbf{13.5\%} & 16.9\% \\
\textbf{Gemini Live 2.5}       & 0.786 & 0.593 & 0.554 & 0.490 & 92.0\% & 7.26s & 14.1\% & \textbf{8.9\%} \\
\textbf{Gemini Live 3.1}       & 0.817 & 0.588 & 0.718 & 0.540 & 78.0\% & \textbf{4.25s} & 19.2\% & 31.7\% \\
\textbf{Grok}         & 0.797 & 0.542 & 0.617 & 0.430 & 94.0\% & 6.65s & 25.5\% & 44.3\% \\
\textbf{Ultravox}         & 0.794 & 0.513 & 0.510 & 0.410 & 96.0\% & 8.40s & 47.9\% & 88.0\% \\
\midrule
\textbf{Cascaded}         & 0.803 & 0.562 & 0.600 & 0.450 & \textbf{100.0\%} & 10.12s & 33.0\% & 26.9\% \\
\bottomrule
\end{tabular}%
}
\caption{Overall Performance and Turn-Taking Metrics. Best value in each column is in \textbf{bold}. Lower latency and lower filler rate are better. Filler denotes the percentage of responses containing filler sentences before the key information.}
\label{tab:all}
\end{table*}

\subsection{Evaluation Metrics}
We score agents across four dimensions that isolate logical reasoning from acoustic responsiveness.

\begin{itemize}
    \item \textbf{Tool Selection F1:} The F1 score over expected vs.\ actual tool calls, penalizing both missed calls (low recall) and hallucinated ones (low precision).

    \item \textbf{Argument Accuracy:} Semantic correctness of generated arguments, judged by GPT-4o to accommodate valid input variations such as date formats, abbreviations, and dynamic variables from prior turns (Appendix~\ref{sec:eval_prompts_arg}).

    \item \textbf{Task Completion (Pass@1):} A binary metric requiring all three conditions simultaneously: the agent invokes exactly the expected tools, and scores perfect argument accuracy for every call. Any single failure yields a fail.

    \item \textbf{Response Quality:} GPT-4o judges whether the agent's spoken transcript accurately fulfills the user's intent in natural language, penalizing correct tool execution that fails to relay results effectively (Appendix~\ref{sec:eval_prompts_response}).

    \item \textbf{Turn-Taking and Latency Dynamics:} From time-aligned execution logs, the \textbf{Turn-take rate} measures the fraction of turns with a natural-timing response. Base latency is $\Delta t = t_{\text{agent\_start}} - t_{\text{user\_end}}$; if $\Delta t < 0$, the event is an \textbf{Interruption}.

    We decompose latency into three components:
    \begin{enumerate}
        \item \textit{First Response Latency:} Time until any speech, including filler sentences (e.g., ``Let me check on that.'').
        \item \textit{Tool Call Latency:} Time until the first API invocation.
        \item \textit{Task Completion Latency:} Time until the agent delivers the factual answer. GPT-4o analyzes ASR chunks to isolate the informational sentence from preceding filler speech (Appendix~\ref{sec:eval_prompts_latency}).
    \end{enumerate}
    We also report the \textbf{Filler Rate}: the fraction of scenarios where the agent emits a content-free filler sentence (e.g., ``Sure, let me look that up'') before the substantive response---a strategy that reduces perceived latency at the potential cost of interrupting users who are still speaking.
\end{itemize}

\section{Results}

\begin{table*}[htbp]
\centering

\small
\begin{tabular}{lccccc}
\toprule
\textbf{Model} & \textbf{Filler} & \textbf{Pause} & \textbf{Hesitation} & \textbf{False Start} & \textbf{Self-Corr} \\
\midrule
\textbf{GPT-Realtime} & \textbf{0.621} & \textbf{0.556} & \textbf{0.700} & \textbf{0.667} & \textbf{0.588} \\
\textbf{Gemini Live 2.5} & \textbf{0.621} & 0.444 & 0.600 & 0.417 & 0.471 \\
\textbf{Gemini Live 3.1} & 0.586 & 0.500 & 0.600 & 0.583 & 0.353 \\
\textbf{Grok} & 0.483 & 0.333 & 0.500 & 0.583 & 0.294 \\
\textbf{Ultravox} & 0.414 & 0.333 & 0.500 & 0.250 & 0.353 \\
\midrule
\textbf{Cascaded} & 0.448 & 0.444 & 0.600 & 0.500 & 0.176 \\
\bottomrule
\end{tabular}
\caption{Robustness to disfluency features (Pass@1 per category).}
\label{tab:disfluency}
\end{table*}

\subsection{Overall Performance}

Table~\ref{tab:all} summarises aggregate scores. GPT-Realtime is the strongest overall performer, leading on tool selection F1 (0.876), argument accuracy (0.680), response quality (0.792), Pass@1 (0.600), and the lowest interruption rate (13.5\%).

Gemini Live 3.1 ranks second on Pass@1 (0.540) with the fastest latency (4.25~s), but its turn-take rate (78.0\%) is the lowest---22 scenarios received no response. Gemini Live 2.5 is more conservative: higher turn-take rate (92.0\%) and low interruption rate (14.1\%), but lower accuracy overall.

Grok and Ultravox occupy the lower accuracy tier (Pass@1 0.430 and 0.410). Ultravox ties GPT-Realtime on turn-take rate (96.0\%) but has the highest interruption rate (47.9\%) and filler rate (88.0\%), indicating it frequently speaks over users with content-free utterances. Grok balances moderate latency (6.65~s) with a 25.5\% interruption rate.

The Cascaded baseline (Pass@1 0.450) is the only system with a perfect turn-take rate (100\%), as its sequential pipeline guarantees a response for every input. This reliability comes at the cost of the highest latency (10.12~s) and a 33.0\% interruption rate.

\subsection{Performance by Difficulty}

Table~\ref{tab:difficulty} breaks down Pass@1 by difficulty. GPT-Realtime leads at every level (Easy 0.750, Medium 0.588, Hard 0.433). Performance degrades consistently with complexity across all systems. The Cascaded baseline scores competitively on Easy tasks (0.639)---benefiting from GPT-4o's single-step reasoning---but degrades sharply on Hard (0.233), suggesting that ASR error propagation compounds with task complexity. Grok shows the steepest decline (0.200 on Hard), confirming that multi-step reasoning under disfluent speech is a key differentiator.

\begin{table}[htbp]
\centering
\small
\begin{tabular}{lccc}
\toprule
\textbf{Model} & \textbf{Easy} & \textbf{Medium} & \textbf{Hard} \\
\midrule
\textbf{GPT-Realtime} & \textbf{0.750} & \textbf{0.588} & \textbf{0.433} \\
\textbf{Gemini Live 2.5} & 0.667 & 0.500 & 0.267 \\
\textbf{Gemini Live 3.1} & 0.694 & \textbf{0.588} & 0.300 \\
\textbf{Grok} & 0.583 & 0.471 & 0.200 \\
\textbf{Ultravox} & 0.556 & 0.382 & 0.267 \\
\midrule
\textbf{Cascaded} & 0.639 & 0.441 & 0.233 \\
\bottomrule
\end{tabular}
\caption{Pass@1 Performance by scenario difficulty.}
\label{tab:difficulty}
\end{table}

\begin{table*}[htbp]
\centering

\small
\begin{tabular}{lcccc}
\toprule
\textbf{Model} & \textbf{Ecommerce} & \textbf{Finance} & \textbf{Housing} & \textbf{Travel} \\
\midrule
\textbf{GPT-Realtime}     & \textbf{0.552} & \textbf{0.960} & \textbf{0.308} & \textbf{0.600} \\
\textbf{Gemini Live 2.5}     & 0.483 & 0.760 & 0.231 & 0.500 \\
\textbf{Gemini Live 3.1} & 0.448 & 0.920 & 0.269 & 0.550 \\
\textbf{Grok}        & 0.414 & 0.760 & 0.115 & 0.450 \\
\textbf{Ultravox}   & 0.345 & 0.680 & 0.192 & 0.450 \\
\midrule
\textbf{Cascaded}   & 0.414 & 0.800 & 0.192 & 0.400 \\
\bottomrule
\end{tabular}
\caption{Pass@1 Performance breakdown by task domain. Best value in each column is shown in \textbf{bold}.}
\label{tab:domain_performance}
\end{table*}

\subsection{Robustness to Disfluency}
Table~\ref{tab:disfluency} conditions Pass@1 on disfluency type. GPT-Realtime leads or ties on every category, with a particularly strong self-correction score (0.588) that substantially exceeds all other systems. Gemini Live 2.5 is second on self-corrections (0.471) but lags on false starts (0.417); Gemini Live 3.1 is more balanced but trails on self-corrections (0.353), suggesting the version update improved general robustness without improving state rollback. Pauses are the weakest category across most systems (Grok and Ultravox both at 0.333), highlighting a shared difficulty in detecting whether a user has finished speaking.

\subsection{Performance by Domain}
Table~\ref{tab:domain_performance} breaks down Pass@1 by domain. Finance is the easiest domain for all models (GPT-Realtime 0.960, Gemini Live 3.1 0.920); Housing is the hardest (GPT-Realtime leads at 0.308, Grok at 0.115). GPT-Realtime leads on all four domains. Ultravox shows notable domain imbalance: its E-commerce (0.345) and Housing (0.192) scores are among the lowest, pointing to weaknesses in multi-entity order handling and complex constraint reasoning.

\subsection{Latency Analysis}

Table~\ref{tab:latency} decomposes latency into first word, tool call, and task completion. Gemini Live 3.1 is fastest on all three (3.95~s / 2.21~s / 4.25~s), though this speed likely contributes to its low turn-take rate. GPT-Realtime maintains moderate latency (6.89~s) alongside the best accuracy. Grok is competitive (6.65~s) with the fastest first-word response among non-Gemini systems.

Ultravox presents a distinctive \emph{inverted} latency profile: its first-word latency (3.88~s) is among the fastest, yet its tool-call latency (6.01~s) is the slowest---the model speaks \emph{before} it calls any tool. This is explained by its 88.0\% filler rate: Ultravox almost always emits a filler sentence (e.g., ``Let me check on that'') before initiating the API call. While this reduces perceived wait time, filler speech frequently overlaps with the user's utterance (47.9\% interruption rate), and deferring tool execution until after the filler inflates task-completion latency to 8.40~s.

The Cascaded system is slowest overall (10.12~s), with first-word delay (8.78~s) dominating---confirming that the sequential Whisper$\rightarrow$LLM$\rightarrow$TTS chain creates a bottleneck that end-to-end models avoid through concurrent processing.

\begin{table}[htbp]
\centering
\resizebox{\columnwidth}{!}{%
\begin{tabular}{lccc}
\toprule
\textbf{Model} & \textbf{First Word} & \textbf{Tool Call} & \textbf{Task Compl.} \\
\midrule
\textbf{GPT-Realtime}     & 6.36 & 3.89 & 6.89 \\
\textbf{Gemini Live 2.5}     & 7.03 & 4.61 & 7.26 \\
\textbf{Gemini Live 3.1}     & \textbf{3.95} & \textbf{2.21} & \textbf{4.25} \\
\textbf{Grok}       & 5.97 & 0.63 & 6.65 \\
\textbf{Ultravox}       & 3.88 & 6.01 & 8.40 \\
\midrule
\textbf{Cascaded}       & 8.78 & 3.15 & 10.12 \\
\bottomrule
\end{tabular}%
}
\caption{Mean latency breakdown in seconds. Lower is better.}
\label{tab:latency}
\end{table}

\section{Discussion}

\subsection{Turn-Taking Trade-offs}

How a model handles silence is as important as whether it selects the right tool. The Cascaded baseline achieves 100\% turn-take rate by design, but among end-to-end models the picture is more nuanced. Ultravox ties GPT-Realtime at 96.0\% turn-take rate but interrupts users 47.9\% of the time---nearly half of all turns. GPT-Realtime strikes the best balance: 96.0\% turn-take with only 13.5\% interruption, implying well-calibrated voice activity detection. Grok occupies the middle ground (94.0\% / 25.5\%).

\subsection{Pre-emptive Tool Use and Interruption Patterns}

We define a \textit{pre-emptive tool call} as one invoked before the user finishes speaking (negative tool-call latency). Pre-emptive rates vary widely: Grok 41.6\%, Ultravox 23.2\%, Gemini Live 2.5 17.9\%, Gemini Live 3.1 16.9\%, GPT-Realtime 10.8\%.

Pre-emptive tool-call rates do not predict interruption rates. Grok has the highest pre-emptive rate (41.6\%) but only moderate interruptions (25.5\%)---it processes tools silently while letting the user finish. Ultravox has a lower pre-emptive rate (23.2\%) but the highest interruption rate (47.9\%), confirming that its interruptions stem from premature \textit{speech}, not premature \textit{tool invocation}. This reveals two distinct failure modes: \textit{silent pre-processing} (Grok), where reasoning runs early but speech is deferred, versus \textit{eager speaking} (Ultravox), where the model speaks before the user has finished.

\subsection{Self-Corrections Remain Difficult}

All systems handle surface-level disfluencies (fillers, hesitations) reasonably well but struggle when a user changes intent mid-utterance. Even GPT-Realtime, the leader at 0.588, fails on over 40\% of self-correction scenarios. Gemini Live 2.5 follows at 0.471; Gemini Live 3.1 (0.353) and Ultravox (0.353) perform worse despite general capability improvements; Grok is lowest (0.294); and the Cascaded system scores just 0.176. The core challenge is that models commit intermediate parameters before the correction arrives, and reliable rollback requires distinguishing provisionally set values from explicitly confirmed ones.

\subsection{Gemini Live 3.1: The Silent Worker}

Gemini Live 3.1 exhibits the most striking behavioral pattern. Despite the fastest latency (4.25~s) and competitive accuracy when it responds, it produces no speech in 22\% examples (78.0\% turn-take rate). Crucially, \textbf{86\% silent cases still executed tool calls}---the model identified and invoked APIs but never generated speech. Three of these achieved perfect tool selection and argument accuracy, costing a potential Pass@1 improvement.

This ``silent worker'' phenomenon concentrates in harder scenarios: 0\% of easy, 23.5\% of medium, and 46.7\% of hard scenarios received no response. The failure mode is a disconnect between reasoning and speech generation---architecturally distinct from other models, where no-response cases correspond to complete processing failure (e.g., all 4 of GPT-Realtime's silent cases had zero tool calls).

\subsection{Cascaded Pipeline: Reliable but Slow}

The Cascaded baseline (Whisper $\rightarrow$ GPT-4o $\rightarrow$ OpenAI TTS) isolates the cost of the traditional modular architecture. Although it shares the same underlying LLM as GPT-Realtime, its Pass@1 is markedly lower (0.450 vs.\ 0.600), indicating that ASR-introduced errors propagate downstream. The gap is starkest on self-corrections: Cascaded scores only 0.176---the lowest of all systems---versus GPT-Realtime's 0.588. Because Whisper may finalize the original (incorrect) transcription before the user's correction arrives, the downstream LLM has no opportunity for state rollback.

Conversely, the pipeline guarantees engagement: its turn-take rate is a perfect 100\%, eliminating the ``silent worker'' failures seen in Gemini Live 3.1. This reliability comes at the cost of the highest task-completion latency (10.12~s, $\sim$2.4$\times$ Gemini Live 3.1), dominated by the first-word delay (8.78~s). The sequential Whisper$\rightarrow$LLM$\rightarrow$TTS chain creates an irreducible bottleneck that end-to-end models sidestep through concurrent processing, quantifying the core trade-off between modular reliability and native-speech-model speed.

\subsection{Qualitative Case Studies}

To illustrate the interplay between accuracy, latency, and turn-taking behavior, we examine two representative hard-difficulty scenarios in detail, comparing all six systems on tool correctness, response timing, and filler usage.

\paragraph{Case 1: Multi-Step Chain Without Disfluency.}
Table~\ref{tab:case_finance18} shows \texttt{finance\_18}, a three-tool chain: convert 1{,}000~USD to EUR, update credit-card autopay to savings, and check premium-card benefits. No disfluency is present---the user issues all three requests clearly in a single utterance.

\begin{table}[htbp]
\centering
\resizebox{\columnwidth}{!}{%
\begin{tabular}{lcccccc}
\toprule
\textbf{Model} & \textbf{Tool} & \textbf{Arg} & \textbf{Resp} & \textbf{1\textsuperscript{st} Resp} & \textbf{Tool Call} & \textbf{Compl.} \\
 & \textbf{Sel} & \textbf{Acc} & \textbf{Qual} & \textbf{(s)} & \textbf{(s)} & \textbf{(s)} \\
\midrule
\textbf{GPT-RT}  & 1.00 & 1.00 & 1.00 & 8.48 & 6.40 & 9.20 \\
\textbf{Gem 2.5} & 1.00 & 1.00 & 1.00 & 6.16 & 3.60 & 6.16 \\
\textbf{Gem 3.1} & 1.00 & 1.00 & 1.00 & \textbf{3.36} & \textbf{2.43} & \textbf{3.92} \\
\textbf{Grok}    & 1.00 & 1.00 & 1.00 & 5.20 & 2.39 & 5.92 \\
\textbf{Ultravox}& 1.00 & 0.67 & 0.00 & 2.64 & 6.31 & 8.16 \\
\textbf{Cascaded}& 1.00 & 0.67 & 0.00 & $-$2.40 & --- & 8.64 \\
\bottomrule
\end{tabular}%
}
\caption{Case study: \texttt{finance\_18} (hard, 3 tools, no disfluency). All four end-to-end models achieve perfect accuracy; \textbf{Gemini 3.1 completes in 3.92\,s vs.\ GPT-Realtime's 9.20\,s} (2.3$\times$ faster).}
\label{tab:case_finance18}
\end{table}

Four of the six systems---GPT-Realtime, Gemini Live~2.5, Gemini 3.1, and Grok---achieve perfect scores on all three metrics. Gemini Live~3.1 stands out: it issues the first tool call in 2.43~s and completes the task in 3.92~s, 2.3$\times$ faster than GPT-Realtime (9.20~s). Both use a brief filler (GPT: ``All set.''; Gemini~3.1: ``Sure,'') to cover the tool-execution gap, but GPT's filler arrives at 8.48~s while Gemini~3.1's arrives at 3.36~s. Ultravox selects all three tools correctly but converts USD to \textit{NGN} (Nigerian Naira) instead of EUR, illustrating an ASR or reasoning error under otherwise clean input. The Cascaded pipeline similarly misroutes the currency (USD$\rightarrow$USD) and begins speaking \textit{before} the user finishes ($-$2.40~s first response).

\paragraph{Case 2: Double Self-Correction Under Disfluency.}
Table~\ref{tab:case_travel19} presents \texttt{travel\_19}, where the user corrects both destination (Rome$\rightarrow$Milan) and date (June~1$\rightarrow$June~3) mid-utterance.

\begin{table}[htbp]
\centering
\resizebox{\columnwidth}{!}{%
\begin{tabular}{lcccccc}
\toprule
\textbf{Model} & \textbf{Tool} & \textbf{Arg} & \textbf{Resp} & \textbf{1\textsuperscript{st} Resp} & \textbf{Tool Call} & \textbf{Compl.}  \\
 & \textbf{Sel} & \textbf{Acc} & \textbf{Qual} & \textbf{(s)} & \textbf{(s)} & \textbf{(s)} \\
\midrule
\textbf{GPT-RT}  & 1.00 & 1.00 & 1.00 & 4.40 & 2.55 & 4.40  \\
\textbf{Gem 2.5} & 1.00 & 1.00 & 0.00 & 5.28 & 4.23 & 5.28  \\
\textbf{Gem 3.1} & 1.00 & 0.00 & 0.00 & 2.56 & $-$2.27 & 2.56  \\
\textbf{Grok}    & 0.50 & 0.00 & 1.00 & --- & $-$2.47 & --- \\
\textbf{Ultravox}& 1.00 & 1.00 & 0.00 & $-$1.36 & $-$0.18 & 3.04  \\
\textbf{Cascaded}& 0.67 & 0.00 & 0.00 & 8.08 & 1.79 & 8.08  \\
\bottomrule
\end{tabular}%
}
\caption{Case study: \texttt{travel\_19} (hard, self-correction, state rollback). Only \textbf{GPT-Realtime correctly applies both corrections} (Rome$\rightarrow$Milan, June~1$\rightarrow$3). Gemini~3.1's pre-emptive tool call ($-$2.27\,s) locks in the stale destination.}
\label{tab:case_travel19}
\end{table}

Only GPT-Realtime achieves a perfect score, correctly searching for flights to \textit{Milan} on \textit{June~3}. Gemini Live~3.1's speed advantage becomes a liability: its tool-call latency of $-$2.27~s means the API was invoked \textit{before the user finished correcting}, locking in \texttt{destination=``Rome''} (the original, uncorrected value). The Cascaded pipeline also uses ``Rome,'' but for a different reason: Whisper finalizes the initial transcription before the correction arrives, so the downstream LLM never receives the updated intent. Ultravox correctly resolves Milan but begins speaking 1.36~s before the user finishes (filler: ``I'll update your passport number right away''), interrupting the user with an unrelated statement.

Taken together, the two cases reveal a fundamental tension: Gemini Live~3.1's concurrent processing enables 2$\times$ faster task completion on straightforward multi-step chains (Case~1), but the same pre-emptive mechanism prevents state rollback when users change their minds mid-utterance (Case~2). Designing when to commit tool parameters---eagerly for speed or conservatively for correctness---remains an open challenge for real-time voice agents.

\section{Conclusion}
In this work, we introduce Full-Duplex-Bench-v3, the first benchmark to evaluate real-time voice agents on multi-step tool execution using natural, unscripted human speech. Our evaluation of six leading models reveals a clear trade-off between response speed, conversational flow, and reliable reasoning. While end-to-end models are significantly faster than traditional cascaded pipelines, their design introduces new challenges. For example, optimizing for minimal delay leads to diverse turn-taking behaviors—ranging from silent background processing (as seen in Gemini Live 3.1's "silent worker" pattern) to the eager use of fillers that can cause unintended interruptions (like Ultravox).

Most importantly, FDB-v3 shows that handling mid-sentence corrections remains an open challenge for all current models. The same early processing that makes these agents fast frequently locks in outdated user intents, preventing even top models like GPT-Realtime from successfully updating their actions on the fly. Ultimately, our findings suggest that the next frontier for voice agents is not just reducing latency. Instead, future architectures must balance fast tool execution with the flexibility to handle the unpredictable and constantly changing nature of real human conversation.

\section*{Limitations}
All cloud-based model evaluations were executed from a single fixed server region with high-bandwidth connections to ensure fair latency comparisons. Nevertheless, measured latencies for proprietary models inherently include non-deterministic network overhead and varying server-side loads.

Our zero-latency local mock APIs isolate model reasoning from external confounds but do not test robustness to real-world network anomalies such as API timeouts, access denials, or malformed responses.


\bibliography{custom}

\appendix

\section{Evaluation Prompts}
\label{sec:eval_prompts}

We detail the exact prompts used in our evaluation pipeline. All LLM-based evaluations use GPT-4o with structured JSON output.

\subsection{Argument Accuracy Judge}
\label{sec:eval_prompts_arg}
This prompt is used to evaluate whether the agent passed semantically correct arguments to each tool call (\S4, Argument Accuracy).

\begin{tcolorbox}[colback=green!3!white, colframe=green!50!black, title=Argument Accuracy Judge Prompt, fonttitle=\bfseries\small, breakable, sharp corners, boxrule=0.5pt, left=3pt, right=3pt, top=3pt, bottom=3pt]
{\footnotesize\ttfamily
You are evaluating whether an AI voice agent called a function with correct arguments.\par\medskip
Function: \{function\_name\}\par
Expected arguments: \{expected\_args\}\par
Actual arguments: \{actual\_args\}\par\medskip
Rules:\par
1. Arguments that start with ``\$'' (like ``\$RESULT\_0.flights[0].flight\_id'') are dynamic references --- the actual value should be any real value that could plausibly come from a previous API call.\par
2. Minor formatting differences are fine: ``August 20'' == ``2026-08-20'', ``New York'' == ``new york''.\par
3. ``Las Vegas'' == ``Vegas'' --- abbreviations and common aliases are acceptable.\par
4. Numeric tolerance: $\pm$5\% is acceptable.\par
5. doc\_type: ``driver\_license'' == ``driver license'' (underscore vs space).\par\medskip
Respond with ONLY a JSON object:\par
\{"correct": true/false, "explanation": "brief reason"\}
}
\end{tcolorbox}

\subsection{Response Quality Judge}
\label{sec:eval_prompts_response}
This prompt evaluates whether the agent's spoken response correctly fulfills the user's intent (\S4, Response Quality).

\begin{tcolorbox}[colback=orange!3!white, colframe=orange!50!black, title=Response Quality Judge Prompt, fonttitle=\bfseries\small, breakable, sharp corners, boxrule=0.5pt, left=3pt, right=3pt, top=3pt, bottom=3pt]
{\footnotesize\ttfamily
You are evaluating whether an AI voice agent successfully completed the user's requested task.\par\medskip
Expected Task/Action: ``\{expected\_intent\}''\par
Actual Agent Spoken Response: ``\{actual\_transcript\}''\par\medskip
Evaluation criteria:\par
1. Did the agent perform the CORRECT actions (right tools, right parameters)?\par
2. Did the response indicate the task was completed or is being handled?\par
3. It is FINE if the agent provides MORE detail than expected (e.g., giving specific results, prices, confirmation numbers). Providing additional helpful information is NOT a penalty.\par
4. It is INCORRECT if the agent says it cannot perform the action, lacks tools, or refuses.\par
5. It is INCORRECT if the agent performs the WRONG action (e.g., wrong destination, wrong document type).\par
6. Partial delivery of multi-step tasks (e.g., completes 2 of 3 required steps) should be scored 0.\par\medskip
Respond with ONLY a JSON object:\par
\{"correct": true/false, "explanation": "brief reason"\}
}
\end{tcolorbox}

\subsection{Key Information Identifier (Latency)}
\label{sec:eval_prompts_latency}
This system prompt is used to decompose agent speech into filler and key information for task completion latency measurement (\S4, Task Completion Latency).

\begin{tcolorbox}[colback=purple!3!white, colframe=purple!50!black, title=Key Information Identifier Prompt, fonttitle=\bfseries\small, breakable, sharp corners, boxrule=0.5pt, left=3pt, right=3pt, top=3pt, bottom=3pt]
{\footnotesize\ttfamily
You are an expert audio transcript analyst for a voice AI assistant evaluation.\par\medskip
You will receive:\par
1. USER\_SPEECH\_END\_REL: timestamp (seconds) when the user finished speaking.\par
2. ASR\_CHUNKS: list of the AI agent's spoken words with [start, end] timestamps.\par
3. TOOL\_CALLS: list of tool calls the agent made (with timestamps).\par\medskip
The typical flow after a user query is:\par
{[}User finishes{]} $\rightarrow$ (silence) $\rightarrow$ {[}Filler sentence{]} $\rightarrow$ (silence during tool execution) $\rightarrow$ {[}Key information response{]}\par\medskip
Your task: Identify and separate the agent's speech into:\par\medskip
1. \textbf{filler\_sentence}: Conversational filler like ``Let me check that for you'', ``Sure, I'll look that up'', ``One moment please''. If the agent immediately starts with the factual answer, this is ``'' (empty).\par\medskip
2. \textbf{key\_info\_sentence}: The part of the response containing the ACTUAL factual information or task confirmation, e.g., ``I found flights to London starting at \$450'' or ``Your passport has been updated with number E772211''. This is the sentence the user is actually waiting for.\par\medskip
3. \textbf{key\_info\_start\_time}: The timestamp (from asr\_chunks) when the key information sentence begins.\par\medskip
Output ONLY a valid JSON object:\par
\{"filler\_sentence": "string or empty", ``filler\_start\_time'': float or null, ``filler\_end\_time'': float or null, ``key\_info\_sentence'': ``string'', ``key\_info\_start\_time'': float, ``key\_info\_end\_time'': float\}
}
\end{tcolorbox}

\end{document}